\documentclass[10pt,aps,notitlepage,twocolumn,floatfix,pra]{revtex4-1}

\usepackage{amsfonts}
\usepackage{amsmath}
\usepackage{amssymb}
\usepackage[dvips]{graphics,graphicx}
\begin{document}

\author{D. Ursrey}
\author{F. Anis}
\author{B. D. Esry}

\affiliation{J.R. Macdonald Laboratory, Kansas State University, Manhattan, Kansas 66506}

\title{Multiphoton Dissociation of $\mathbf{HeH^{+}}$ below the $\mathbf{He^{+}(1s)+H(1s)}$ Threshold} 

\begin{abstract}
We discuss the strong-field dynamics of HeH$^+$, the simplest stable heteronuclear molecule, focusing
on identifying a laser regime for which there is a sufficient dissociation signal for experimental measurement.
We numerically solve the time-dependent Schr\"odinger equation to obtain total dissociation probabilities,
kinetic energy release spectra, and momentum distributions for wavelengths from 800~nm to 2400~nm.
The suitability of this simple system as a prototype for understanding the strong-field nuclear
dynamics of heteronuclear dissociation is discussed.
\end{abstract}

\date{\today}  
\maketitle

\section{Introduction}

There is considerable interest in the dynamics of molecules in intense laser
fields~\cite{0034-4885-67-5-R01,0953-4075-28-3-006}.
Being the simplest molecules, $\mathrm{H_{2}^{+}}$ and $\mathrm{D_{2}^{+}}$ have been at the center of many of these studies.  
Their simplicity allows for theoretical studies with minimal approximations
{\cite{0953-4075-28-3-006,PhysRevA.68.053401,PhysRevA.72.033413,1742-6596-88-1-012046,PhysRevA.77.033416,PhysRevLett.100.133001,0953-4075-42-9-091001,PhysRevLett.103.103006}, while the accuracy of these calculations opens the possibility for quantitative comparison with experimental results~\cite{PhysRevA.68.053401,PhysRevA.72.033413}.
Because of these features, $\mathrm{H_{2}^{+}}$ and $\mathrm{D_{2}^{+}}$ serve as benchmark systems, providing insight into how more complicated molecules behave in intense laser fields.

In these benchmark studies of $\mathrm{H_{2}^{+}}$ and its isotopes, many interesting phenomena have been calculated and measured.  
Among these are above threshold dissociation~\cite{PhysRevLett.64.515}, 
below threshold dissociation~\cite{PhysRevLett.83.3625}, coherent control of dissociation through variation of carrier envelope phase of a single color pulse~\cite{PhysRevLett.93.163601,0953-4075-42-8-085601,Kling14042006} 
or the delay between two pulses with different colors~\cite{PhysRevLett.75.2815,PhysRevLett.74.4799,PhysRevLett.103.223201}, 
and bond softening \cite{PhysRevLett.64.1883,PhysRevLett.70.1077,0953-4075-38-15-L01}.
Given the utility and success of these H$_2^+$ studies, a natural next step is to identify 
and study a similarly fundamental heteronuclear system
Carrying out calculations on such a system will not only provide 
additional insight into how the aforementioned phenomena generalize, 
but it will also allow for the identification of new phenomena that are not present in homonuclear molecules.
We exclude the simplest heteronuclear molecule, $\mathrm{HD^{+}}$, 
because it does not produce an electronic asymmetry within the usual infinite nuclear mass Born-Oppenheimer approximation.

Thus,
we propose $\mathrm{HeH^{+}}$ as a benchmark system for studying the nuclear response of heteronuclear molecules to intense laser fields.
The viability of this species as an experimental target has already been established.
Specifically, ion beams of HeH$^{+}$ have been produced to
demonstrate the existence of metastable HeH$^{2+}$~\cite{HeH2+,HeH2+Lifetime} and, more recently,
to serve as targets for an ultraviolet free-electron laser~\cite{HeH+FEL}.
In this paper, we will focus on dissociation of HeH$^+$ at far- to mid-infrared wavelengths between 800~nm and 2400~nm.
Intense, short pulses with such long wavelengths are becoming increasingly available experimentally
and have been used in recent years to study high harmonic generation~\cite{LongWaveHHG}
and ionization~\cite{LongWaveIon,KlingLongWave}.  We will limit our exploration to 
five and ten cycle laser pulses with intensities between 10$^{12}$~W/cm$^2$ 
and $10^{14}$~$\mathrm{W/cm^{2}}$.
Our results show that although the dissociation probability under these conditions is small for the 
more standard 800~nm pulses, the longer wavelengths in our parameter space are capable of producing 
substantial dissociation, making this process experimentally accessible.

We argue that the large difference in energy between the ground and first excited electronic channels for $\mathrm{HeH^{+}}$ causes permanent dipole transitions to be much more probable than electronic excitation for the wavelengths considered.  
Because the system is dominated by single channel effects, it behaves analogously to an atom in a laser field.  This atom-like behavior allows us to understand physical observables calculated for $\mathrm{HeH^{+}}$ using pictures previously developed to study intense field ionization.  Moreover, it opens up interesting opportunities to observe phenomena usually associated with atoms in a molecular system. 

\begin{figure}[b]
   \includegraphics[scale=0.41]{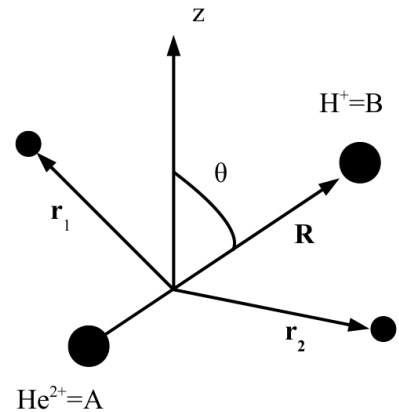}
   \caption{Coordinates used in the Hamiltonian.} 
\end{figure}
\section{Theoretical Methods}

\subsection{Schr\"odinger Equation in the Single Channel Approximation} 

All calculations in this paper use atomic units and are carried out by solving the time-dependent Schr\"odinger equation in the Born-Oppenheimer approximation.
We treat the laser's electric field classically and use the length gauge within the dipole approximation.  
Using the dipole operator 
\begin{align}
\mathbf{d} = -\left( \frac{3 + m_{A}+m_{B}}{2 + m_{A} + m_{B}} \right) \left( \mathbf{r}_{1} + \mathbf{r}_{2} \right) +\left(\frac{m_{A} - 2 m_{B}}{m_{A} + m_{B}}\right) \mathbf{R} \mathrm{,} 
\end{align}
the time-dependent Schr\"odinger equation is given by
\begin{align}
 i\frac{\partial}{\partial t}\Psi  = 
 \left[ -\frac{1}{2\mu_{AB}}\nabla^{2}_{R} + H_{\rm el} 
-\boldsymbol{\cal E}(t) \cdot \mathbf{d} \right] 
 \Psi  ,
 \label{TDSE} 
\end{align}
where $H_{\rm el}$ 
is the electronic part of the Hamiltonian including nuclear repulsion and $\boldsymbol{\cal E}$ is the electric field produced by the incident laser pulse.   
In Eq.~(\ref{TDSE}),
the coordinates are as shown in Fig.~1, and the reduced mass is given by  
\begin{align}
 \frac{1}{\mu_{AB}} = \frac{1}{m_{A}} + \frac{1}{m_{B}} \mathrm{.} 
\end{align}

\begin{figure}
   \includegraphics[width=0.47\linewidth,height=61mm]{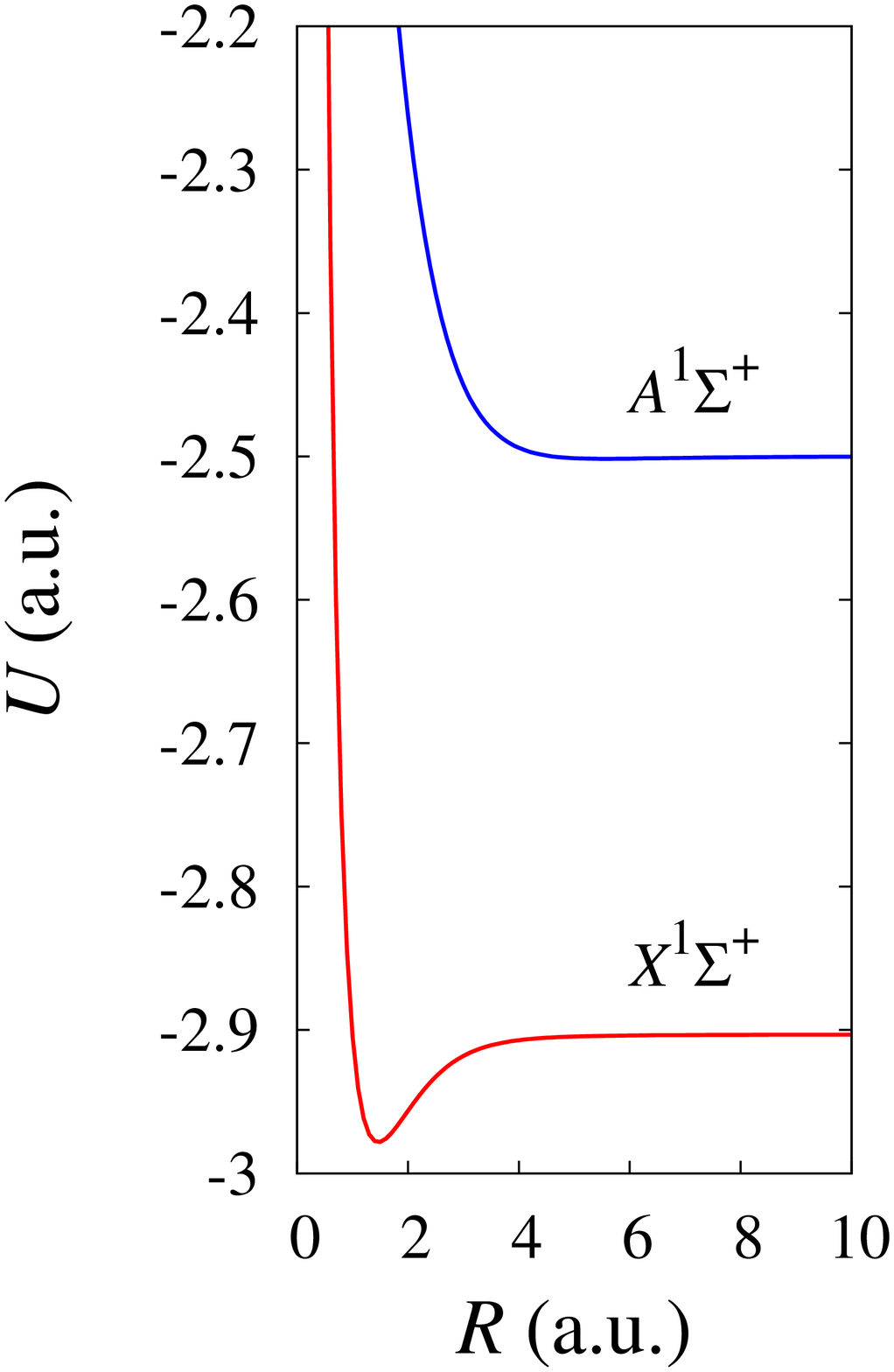} 
   \includegraphics[width=0.47\linewidth,height=61mm]{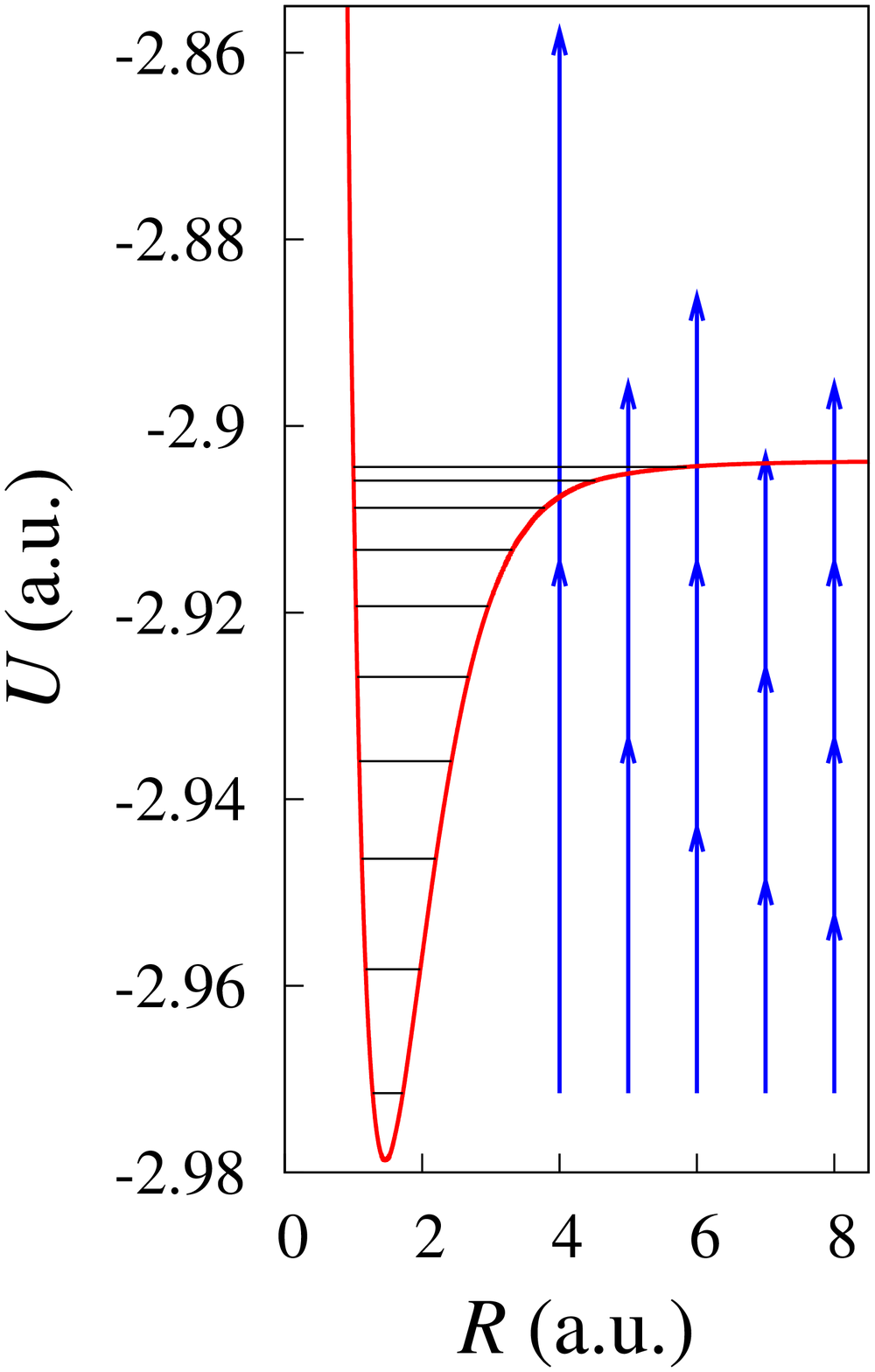}
   \caption{(Color online.) (a) The adiabatic potential energy curves~\cite{0953-4075-43-6-065101} for the ground state of 
        HeH$^+$, $X^1\Sigma^+$, and the first excited singlet state, $A^1\Sigma^+$.
       (b) The $X^1\Sigma^+$ potential used in our calculations~\cite{Kolos1976381} plotted along with the vibrational levels for 
        zero orbital angular momentum.  The arrows indicate the energy after dissociation from the absorption of 
        800~nm, 1200~nm, 1600~nm, 2000~nm, and 2400~nm (from left to right) photons from the vibrational ground state.}
 \label{Potentials}
\end{figure}

The Born-Oppenheimer potentials and electronic wavefunctions are found by solving
\begin{align}
 H_{\mathrm{el}} \Phi_{\nu \Lambda} \left(R\mathit{;}\mathbf{r}_{1} \mathit{,} \mathbf{r}_{2} \right) = U_{\nu \Lambda}\left(R \right) \Phi_{\nu \Lambda} \left(R \mathit{;}\mathbf{r}_{1} \mathit{,} \mathbf{r}_{2} \right) \mathrm{.}
\end{align}
In this paper, we use the potentials from Ref.~\cite{Kolos1976381}.  
Because $\Phi_{\nu \Lambda}(R;{\bf r}_1, {\bf r}_2)$ form a complete 
set in the electronic coordinates, we can write
\begin{equation} 
 \Psi = \sum_{J \alpha \Lambda M} \frac{1}{R} F_{J \alpha \Lambda}(R,t) 
                       \tilde{D}^J_{M \Lambda}(\phi,\theta,0) \Phi_{\alpha \Lambda}(R;{\bf r}_1,{\bf r}_2),
\end{equation}
where $\alpha$ denotes the electronic state, $\Lambda$ is the magnitude of the projection of 
the electronic angular momentum on the internuclear axis, $J$ is the total orbital angular 
momentum, $M$ is the projection of the total orbital angular momentum along the lab frame 
$z$-axis, and $\tilde{D}^J_{M \Lambda}(\phi,\theta,0)$ is the Wigner $D$-function 
normalized over the polar and azimuthal angles $\mathrm{\theta}$ and $\mathrm{\phi}$ that describe  
the orientation of the molecular axis relative to the lab frame.

\begin{figure}
   \includegraphics[width=\columnwidth]{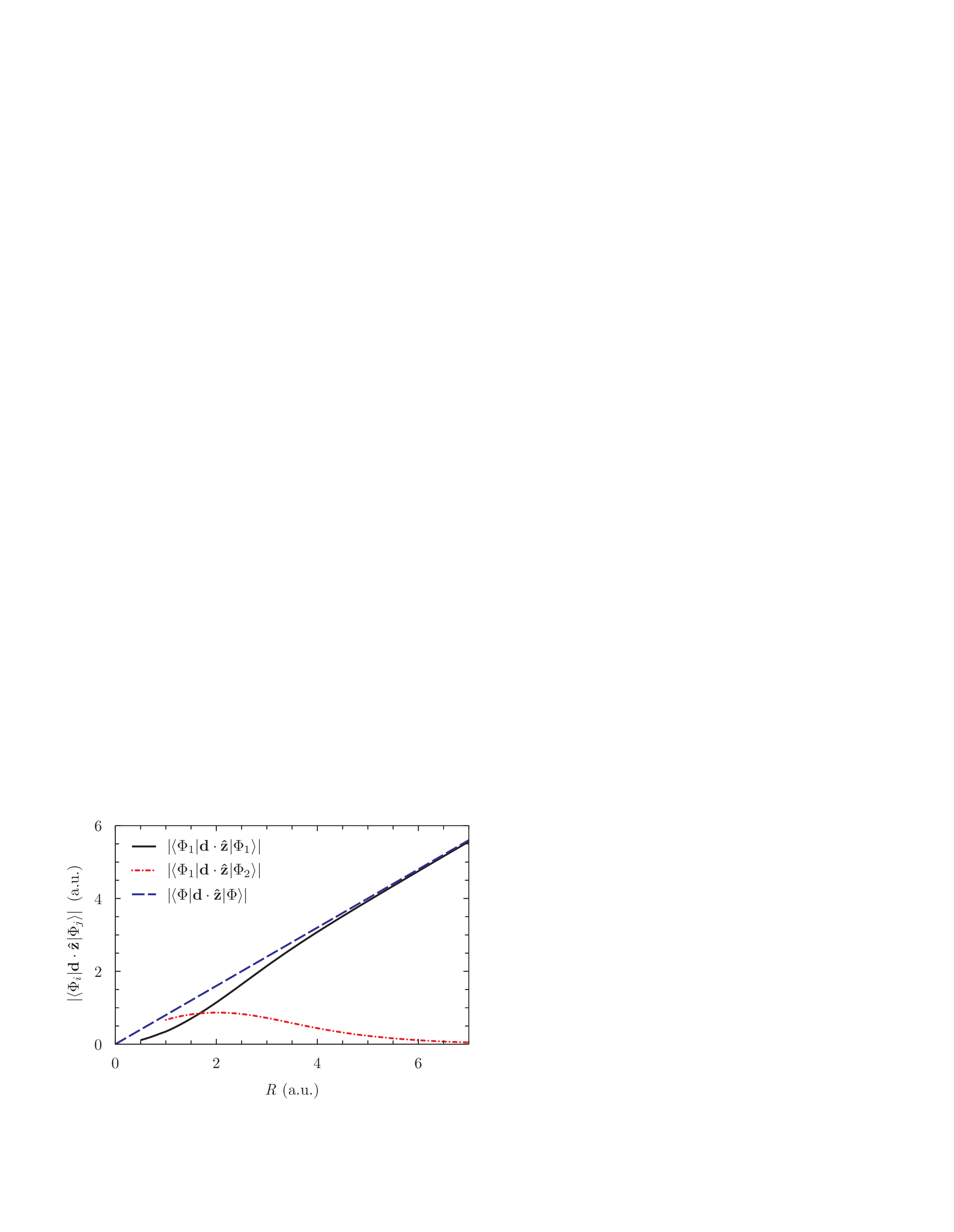} 
   \caption{(Color online.) The body frame $z$-axis projection of the $X^1\Sigma^+$ permanent dipole,
             $\langle \Phi_1 | {\bf d} | \Phi_1 \rangle$, and $X^1\Sigma^+$--$A^1\Sigma^+$ transition dipole,
             $\langle \Phi_1 | {\bf d} | \Phi_2 \rangle$, from Ref.~\cite{0953-4075-43-6-065101}
             along with our approximate permanent dipole $\langle \Phi | \mathbf{d} | \Phi \rangle$ 
             from Eq.~(\ref{ApproxDipole}).} 
  \label{Dipoles}
\end{figure}

In order to reduce the number of electronic channels required in Eq.~(5), we examine the Born-Oppenheimer potentials and dipole couplings to understand the role of electronic excitation for our parameters.  
From Fig.~\ref{Potentials},
we see that there is a difference of at least 0.40 a.u.\
in energy between the ground electronic channel $X^{1}\Sigma^{+}$ and the first excited channel $A^{1}\Sigma^{+}$.
Therefore, a minimum of eight photons is required for electronic excitation with an 800 nm pulse, the shortest wavelength that we consider.
Moreover, we see from Fig.~\ref{Dipoles} that the magnitudes of the permanent dipole, 
$\langle \Phi_1 | \mathbf{d} | \Phi_1 \rangle$, and the $X^1\Sigma^+$--$A^1\Sigma^+$ 
transition dipole matrix element, $\langle \Phi_{1} | \mathbf{d} | \Phi_{2} \rangle$, have 
nearly the same value for 1.0~a.u.\,$\leq R \leq$\,2.0~a.u., and that the permanent dipole is 
larger for $R > 2.0$~a.u. 
In the range where the matrix elements are comparable, however, a minimum of thirteen 
800~nm photons is required for electronic excitation.
Taken together, these considerations suggest that electronic excitation will be negligible for all but very high intensities. 
Therefore, we will neglect electronic excitation in our calculations and keep only the $X^{1}\Sigma^{+}$ state.          
Because our calculations only involve one electronic channel, we will drop the electronic channel labels for the remainder of this work.  

When our calculations were carried out, the extensive results for the dipole matrix elements in Ref.\ \cite{0953-4075-43-6-065101} had not yet been published, so we used an approximate form of the permanent dipole:
\begin{align}
 \left\langle \Phi | \mathbf{d} | \Phi \right\rangle = \frac{m_{A}}{m_{A}+m_{B}} \mathbf{R} \mathrm{.}
 \label{ApproxDipole} 
\end{align}
This form is exact asymptotically as the system dissociates to $\mathrm{He(1s^{2}) + H^{+}}$ and can be derived either classically or by direct evaluation in the separated atom limit. 
Figure~3 compares this approximate dipole to the accurately calculated ones from Ref.~\cite{0953-4075-43-6-065101}.  
The figure shows our approximation to be in good agreement with the calculated dipole with discrepancies for small $R$. 
The fact that the actual dipole is smaller than the one used in our calculations implies that the intensity required to produce a given dissociation probability will be larger than found in our calculations. 
However, lowest order perturbation theory predicts that the actual and reported intensities required to produce the lowest energy kinetic energy release (KER) peak are of the same order of magnitude for the wavelengths considered. 

With the above approximations, Eq.~(5) reduces to
\begin{align}
 \Psi = \sum_{J} \frac{1}{R} F_{J} \left( R,t \right)
 Y_{J0}\left( \phi,\theta \right) \Phi \left(R\mathit{;}\mathbf{r}_{1} \mathit{,} \mathbf{r}_{2} \right) \mathrm{,} 
\end{align}
where $\tilde{D}^{J}_{00} = Y_{J0}$ was used and we have assumed linearly polarized light in order to include only $M=0$ in the expansion.
Substituting this sum into Eq.~(2), projecting out $\Phi Y_{J^{\prime}0}$, and using Eq.~(6) gives the equations we solve for the radial part of the wavefunction: 
\begin{multline} 
 i\frac{\partial}{\partial t} F_{J} =\left[-\frac{1}{2\mu_{AB}}\frac{\partial^{2}}{\partial R^{2}} + \frac{J\left(J+1\right)}{2\mu_{AB}R^{2}} + U \right] F_{J}  \\
  - \sum_{J'}\frac{m_{A}}{m_{A}+m_{B}} {\cal E} \left(t\right) R \sqrt{\frac{4\pi}{3}} \left\langle Y_{J0} | Y_{10} | Y_{J^{\prime}0} \right\rangle F_{J^{\prime}} \mathrm{,}
\end{multline}  
where we have taken the laser polarization in the lab frame $\hat{\mathbf{z}}$ direction.
Equation (8) neglects all coupling between electronic states, including non-Born-Oppenheimer effects 
and Coriolis coupling~\cite{Pack}. 

\subsection{Numerical Methods} 

\begin{figure*}[t]
   \includegraphics[width=\textwidth]{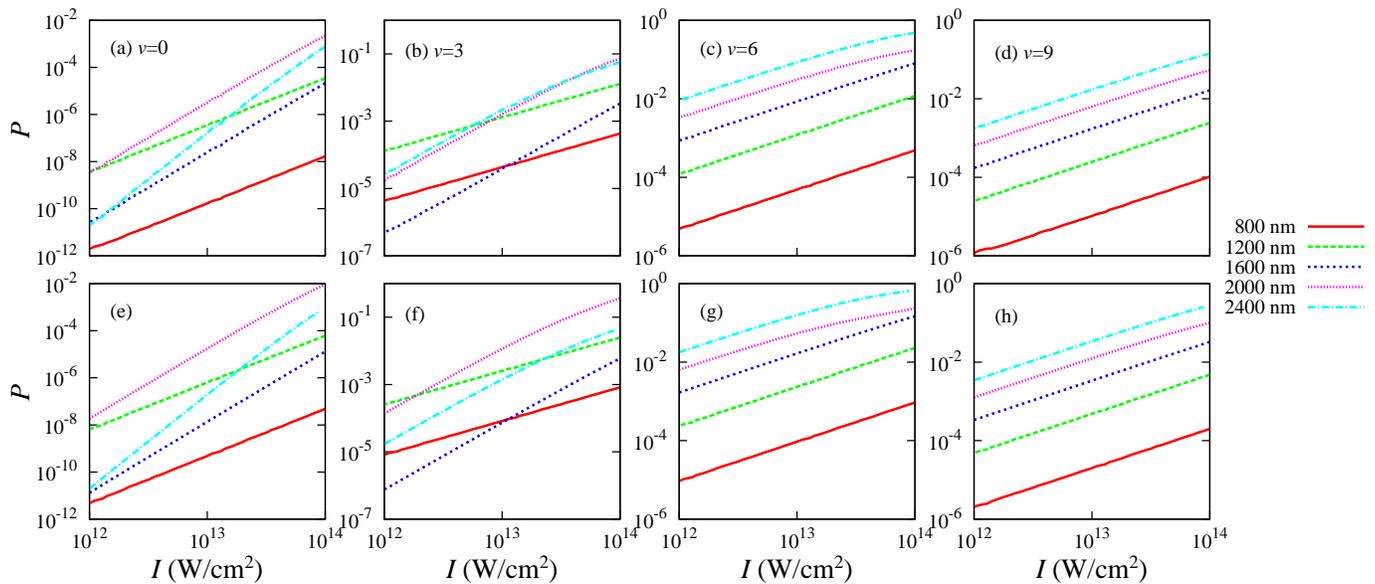}
   \caption{(Color online.) Total dissociation probability $P$ as a function of intensity for different wavelengths, pulse lengths, 
            and initial vibrational states.  
            The top figures are for 5 cycle pulses, while the bottom figures are for 10 cycle pulses.  From left 
            to right, the columns correspond to the initial vibrational states $v$=0, 3, 6, and 9.
            }
   \label{Probs}
\end{figure*}

The remaining task is to solve Eq.~(8).
Time propagation of a given initial state is carried out according to Eq.~(8) using the 
generalized finite differencing scheme from Ref.~\cite{PhysRevA.69.053620} to approximate the 
time-dependent Hamiltonian as the block tridiagonal matrix ${\bf H}(t)$. 
Here, the diagonal blocks are the matrix representation 
of the field-free Hamiltonian, and each off-diagonal block is a diagonal matrix 
with the $J^{\prime}$ to $J^{\prime} \pm 1$ dipole coupling. 
Using $\mathbf{H}\left(t\right)$, time propagation is accomplished for small time steps $\delta$ by
\begin{align}
 \mathbf{F} \left(R,t+\delta  \right) = e^{-i\mathbf{H}\left(t+\delta/2\right)\delta} \mathbf{F} \left(R,t \right) \mathrm{,}
\end{align}
where the components of $\mathbf{F}$ are the $F_{J}$ in Eq.~(7).
The action of this short-time propagator is carried out by applying the split-operator techniques and the Crank-Nicolson method outlined in Ref.~\cite{PhysRevA.74.043411}. 

We use initial vibrational states with $J=0$ and express the electric field as 
\begin{align}
{\cal E}={\cal E}_{0}e^{-\left(t/\tau\right)^{2}} \cos \omega t \mathrm{,}
\end{align}
where $\omega$ is the carrier frequency and $\tau=\tau_{\mathrm{FWHM}}$/$\sqrt{\mathrm{2ln2}}$.
Our calculations are carried out for $\tau_{\mathrm{FWHM}}$ corresponding to both five cycle and ten cycle pulses, using wavelengths in the 800--2400 nm range and intensities between $10^{12}$~$\mathrm{W/cm^{2}}$ and $10^{14}$~$\mathrm{W/cm^{2}}$.
Calculations begin at the time $t_{i}$ when the intensity of the field is $10^{7}$ $\mathrm{W/cm^{2}}$ and end at the time $t_{f}$, after the peak intensity, when the intensity falls off to $10^{8}$ $\mathrm{W/cm^{2}}$.
The time step is $\delta = 0.30$~a.u., and we dynamically increase the number of 
partial waves included in our expansion of $\Psi$ \cite{PhysRevA.72.033413,1742-6596-88-1-012046}, 
leading to a maximal expansion of between 10 and 27 partial waves depending on the parameters of the laser pulse. 
For five cycle pulses, we use a non-uniform radial grid with 
$R_{\mathrm{min}} = 0.20$~a.u., $R_{\mathrm{max}} = 120.0$~a.u.,
and a grid spacing $\Delta R \approx 0.00094 \; \mathrm{a.u.}$ for 
0.20~a.u. $\leq R \leq 8.0$~a.u.
and $\Delta R \approx 0.0094 \; \mathrm{a.u.}$ for $ R > 8.0 \; \mathrm{a.u.}$ 
Three point averaging, where each of three adjacent grid points is weighted equally, is repeatedly applied 20,000 times to all of the grid points in order to smooth the abrupt change in grid spacing 
at $R$=8.0~a.u.  This smoothing is necessary for a non-uniform grid to avoid a loss of accuracy in the radial derivatives of our wavefunction near 
$R=8.0$~a.u.  For ten cycle pulses, we use the same grid spacing but extend $R_{\mathrm{max}}$ to 200.0 $\mathrm{a.u.}$
Convergence with respect to all of these parameters was tested and found to give the KER 
spectrum accurate to three digits up to an energy of 0.20~a.u.

Because part of our aim is to identify the parameters that give the largest dissociation signal, we choose the isotopes of helium and hydrogen that maximize the dipole moment and, thus, the dissociation at a given intensity.  
Equation (6) shows that in order to maximize the dipole, we want the isotopes with the most asymmetric masses.
Consequently, we consider $\mathrm{^{4}He}$ and $\mathrm{^{1}H}$.  For their masses, we use $m_{A} = 7351.67$~a.u. and $m_{B}=1836.15$~a.u.  

\section{Results and Discussion}

\subsection{Dissociation Probability}

\begin{figure}[t]
   \includegraphics[width=0.85\linewidth]{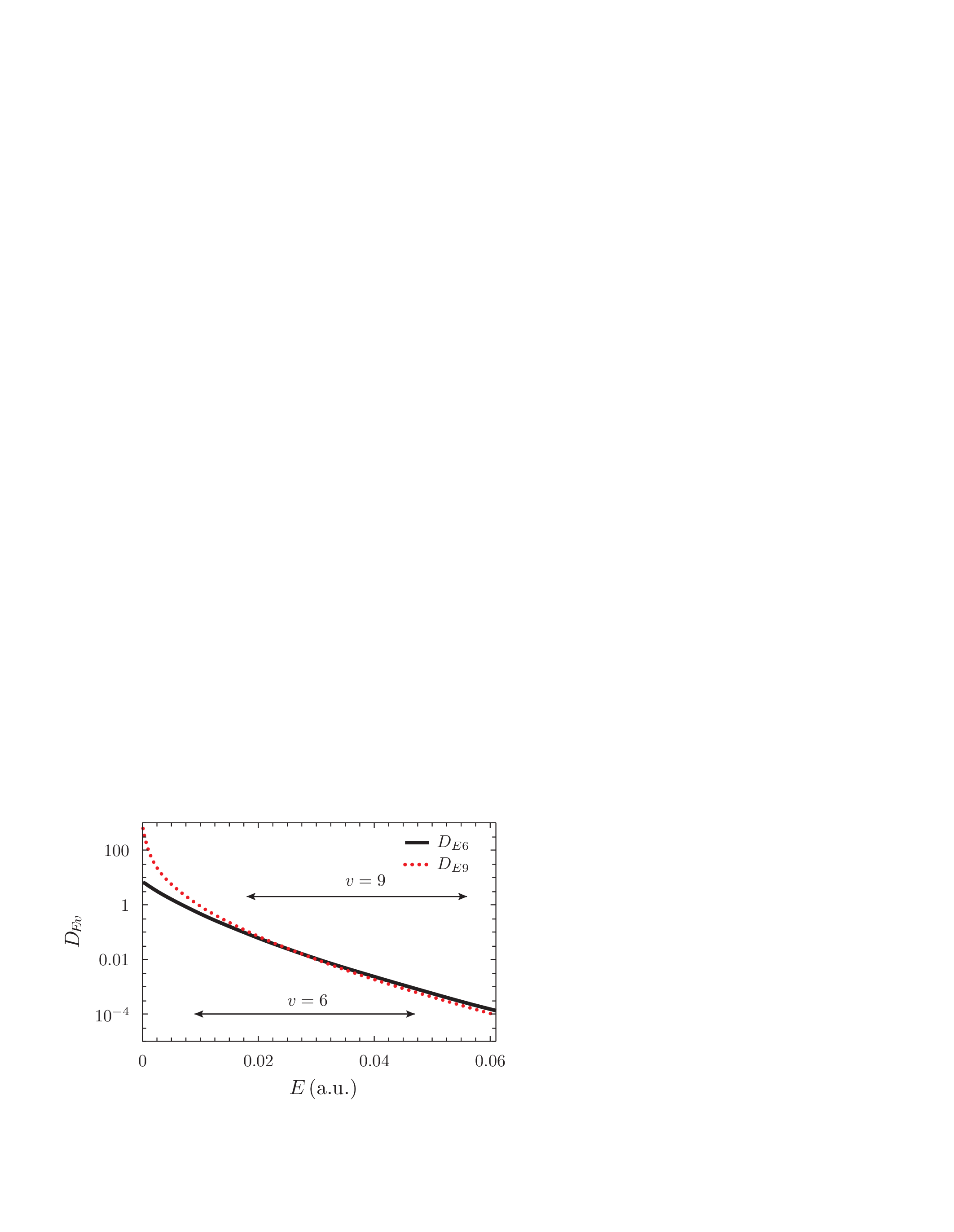} 
   \caption{(Color online.) Bound-free dipole matrix elements, Eq.~(12), as a function of KER $E$ for $v=6$ and $v=9$.  
       The horizontal arrows represent the range of energies expected for one photon absorption for wavelengths 
       between 2400~nm and 800~nm.}
   \label{DipMatElem}
\end{figure}

To quickly locate the parameters that produce substantial dissociation, 
we first consider the total dissociation probability
\begin{equation}
 P = 1 - \sum_{vJ} \big|\langle \chi_{vJ} |  F_{J} (t_{f})\rangle \big|^{2} ,
\end{equation}
for a series of initial vibrational states.
In Eq.~(11), $\chi_{vJ}$ are the field-free ro-vibrational nuclear eigenstates and 
$F_{J} (t_{f})$ 
are the components given by Eq.~(8).
The intensity dependence of $P$ for several initial vibrational states is presented in Fig.~4. 
From the figure, we see that $P \propto I^{n}$ for most initial vibrational states,
which is the result expected from lowest order perturbation theory for \textit{n} photon absorption.   
Deviations from $I^{n}$ do appear for high intensities, indicating that lowest order perturbation theory is no longer adequate.
Figure~2 shows that for an initial $v=0$ state, two photons are required for dissociation at 800~nm and 1200~nm, 
three photons are required at 1600~nm and 2000~nm, and four photons are required at 2400~nm. 
These values of \textit{n} are consistent with the slopes found in Fig.~4.
This same analysis can be applied to explain the slopes seen for other initial states in the figure.    

From Fig.~4, we see that there do exist parameters that produce substantial dissociation: up to 
8.5\% (five cycles) and 16\%
(ten cycles) for $\mathrm{10^{13}}$~W/cm$^{2}$, and up to 48\% and 65\%
for $\mathrm{10^{14}}$~W/cm$^{2}$, respectively.
In all cases, these maximal $P$
occurred for $v=6$ at 2400~nm --- not surprising considering only one photon is required for dissociation from this state.
Dissociation from $v=9$, however, also requires only a single photon, yet 
its dissociation probability is an order of magnitude smaller than for $v=6$.
The difference in dissociation probability for these two states is explained by their bound-free dipole matrix element, 
\begin{align}
 D_{Ev}=\left\langle E, J=1 \bigg| \frac{m_{A}}{m_{A}+m_{B}} R \bigg| \chi_{v0} \right\rangle \mathrm{,}
\end{align}
where $| E, J \rangle$ is the the energy normalized field-free continuum state with 
energy $E$ and angular momentum $J$.
Figure~\ref{DipMatElem} shows $D_{E9}$ and $D_{E6}$ together with the ranges of KER expected in
each case for the one-photon peak for wavelengths between 2400~nm and 800~nm.  Keeping in mind
that the vertical scale in the figure is logarithmic, it is clear that $D_{E9}$ is considerably smaller
than $D_{E6}$ at any given wavelength, yielding, in turn, a much smaller dissociation probability.

The wavelength dependence of $P$
evident in Fig.~\ref{Probs} can also be explained by $D_{Ev}$.
In Fig.~\ref{DipMatElem}, we see that for $v$=6 and 9 the dipole matrix elements decrease monotonically with increasing KER.
In fact, this behavior is a general feature of $D_{Ev}$ for this system.
Therefore, longer wavelengths are expected to give the largest dissociation probabilities 
since they produce dissociating fragments with the lowest KER. 

There is also substantial dissociation for other initial states and laser parameters. 
Both five and ten cycle pulses with $I=10^{14}$ $\mathrm{W/cm}^{2}$ produce dissociation probabilities greater than 1.0\%
for both $\mathrm{\lambda}=2000$~nm and $\mathrm{\lambda}=2400$~nm in all studied initial vibrational states except $v=0$.
Because the largest dissociation occurs for intensities between $10^{13}$~W/cm$^2$ 
and $10^{14}$~W/cm$^2$ and wavelengths between 2000~nm and 2400~nm, we constrain 
our studies of the KER and momentum distribution to this region of parameter space. 

\subsection{Kinetic Energy Release Spectrum}
Due to the fact that HeH$^{+}$ is a benchmark molecular system,
it is useful to try to understand its behavior in a laser field using the pictures already developed for the benchmark molecule $\mathrm{H^{+}_{2}}$. 
In particular, the dressed potentials of the Floquet representation have proven especially useful  in understanding the kinetic energy release spectrum of $\mathrm{H_{2}^{+}}$ \cite{0034-4885-67-5-R01,PhysRevA.77.033416,0953-4075-42-8-085601}. 
These dressed potentials are obtained by shifting the Born-Oppenheimer potentials by integer multiples of $\omega$, keeping only those states connected by one or more application of the dipole selection rules.
At crossings of the dressed potentials, transitions are much more likely because these crossings correspond to a resonance-like condition.  

Because there is only one electronic channel relevant for our laser conditions, shifting it by multiples of $\omega$ produces parallel potentials with no crossings.
When the appropriate centrifugal barriers are included, there are crossings, but only for such high $J$ and photon number that they are not likely to produce transitions.
Thus, non-resonant transitions must dominate for the laser parameters we consider, reducing the utility of the dressed potential picture in this study.
We note that this failure of the dressed photon picture will be generally true for processes dominated by permanent dipole transitions and is not a feature specific to $\mathrm{HeH^{+}}$.
If we were considering shorter wavelengths, such that transitions to the $A^{1}\Sigma^{+}$ (or higher) state were relevant, then the dressed potential picture would again prove useful. 

It seems, then, that to understand the dissociation of $\mathrm{HeH^{+}}$, we must think about the process differently than for $\mathrm{H_{2}^{+}}$. 
A natural picture to adopt is that for atomic ionization because including only a single channel makes the $\mathrm{HeH^{+}}$ nuclear time-dependent Schr\"odinger equation, Eq.~(8), look just like that of a hydrogenic atom.
Pursuing this analogy, Eq.~(8) shows that the ``electron'' in $\mathrm{HeH^{+}}$ has a charge of $m_{A}/\left(m_{A}+m_{B}\right)$, 
a mass of $\mu_{AB}$, and interacts via a short-range central potential. 
This analogy allows us to interpret our results using the pictures that are used to understand intense field ionization.   

Atomic ionization is primarily understood using one of two pictures: tunneling or multiphoton ionization.  
Which picture is most appropriate is customarily determined by the value of the Keldysh parameter $\gamma=\sqrt{E_{b}/2U_{p}}$,
where $E_{b}$ is the binding energy and $U_{p}$=$q^2{\cal E}^{2}/4\mu \omega^{2}$ is the ponderomotive energy~\cite{Becker200235,Bransden}.
In the case of ionization, the charge $q$ and mass $\mu$ in $U_p$ refer to an electron and are unity in atomic
units; they will differ substantially from unity, though, for our analog system.
For HeH$^+(v$=3,$J$=0) in a 2400~nm laser pulse with peak intensity $I$=8.9$\times$10$^{13}$~W/cm$^2$, 
$\gamma$=4.6, placing the system in the multiphoton regime.
For higher-lying ro-vibrational states, the Keldysh parameter will become smaller at these laser parameters, 
dropping to $\gamma$=0.67 for HeH$^{+}(v$=9,$J$=0).  So, with relatively modest --- and likely achievable ---
changes in the laser parameters, HeH$^+$ can studied over the whole range of physical regimes from
multiphoton to tunneling.
For the remainder of this paper, however, we will restrict our consideration
to HeH$^+(v$=3,$J$=0) in a 2400~nm pulse with $I$=8.9$\times$10$^{13}$~W/cm$^2$ which lies
in the multiphoton regime.
Consequently, we expect to see many of the same phenomena that are seen in the multiphoton ionization of atoms, 
such as above threshold ionization (ATI), in the strong-field dissociation of HeH$^{+}$.

One commonly studied observable in atomic systems that reveals these phenomena is
the photoelectron spectrum.  
In HeH$^+$, the analog is the KER spectrum.  That is, the
probability of the molecule dissociating with a relative nuclear kinetic energy \textit{E}, 
\begin{align}
 \frac{d P}{d E} = \sum_{J} |\left\langle E \mathrm{,}J | F_{J} (t_{f} )\right\rangle |^{2} \mathrm{.}
 \label{KERDistDef} 
\end{align} 
A typical result for the KER spectrum in the parameter space that we are considering is shown in Fig.~6.  
This spectrum clearly shows the characteristic photon-spaced peaks expected for above threshold ionization (ATI) --- except, of course, that this is the nuclear KER spectrum, making it above threshold dissociation (ATD).  
In the atomic case, the energies of the ATI peaks are given by $E_{n} = n \omega - E_{b} - U_{p}$, where \textit{n} is the net number of photons absorbed. 
Because the ponderomotive energy is negligible for the parameters we are considering 
($U_{p}/\omega = 0.040$ for the parameters in Fig.~6 and Fig.~7),
this result simplifies to $E_{n}/{\omega} = E_{b}/\omega + n$. 
For $v=3$ and $\lambda=2400 \; \mathrm{nm}$, we expect the first peak to occur for two photon absorption at 
$E_{n=2}/\omega=0.26$, which is in good agreement with the result in Fig.~6. 

\begin{figure}[tb!]
   \includegraphics[width=0.89\linewidth]{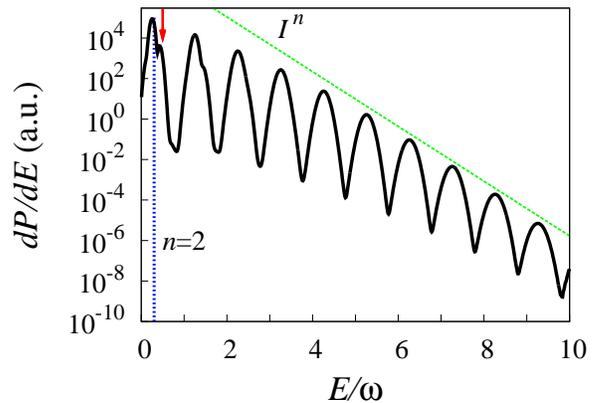} 
   \caption{(Color online.) The KER spectrum for an initial vibrational state of $v=3$ in a five cycle pulse 
         The vertical dashed line at $E/\omega=0.26$
corresponds to the expected location of the first peak, and the dashed diagonal line shows 
the $dP/dE \propto I^{n}$ behavior expected from lowest-order perturbation theory.} 
   \label{KERDistFig}
\end{figure}

It is also important to note the large number of ATD peaks visible in Fig.~6. 
Our result shows more ATD orders than is typical and is a consequence
of the large energy difference between the $X^{1}\Sigma^{+}$ and $A^{1}\Sigma^{+}$ potentials,
which allows for the absorption of many photons before electronic excitation plays a role and blurs the peaks.  
Even so, the highest ATD peak shown in Fig.~6 lies 0.19 a.u. --- more than eleven photons --- below the minimum of the $A^{1}\Sigma^{+}$ potential.

One notable difference between Fig.~6 and a typical ATI spectrum is the absence of the plateau usually seen for $ 3.17 U_{p} < E < 10 U_{p}$ \cite{Becker200235}. 
Because the ponderomotive energy is small relative to the frequency of the laser, our ATD plateau would occur between 
$E/\omega = 0.13$ and $E/\omega = 0.40$.
The fact that $10U_{p}/\omega - 3.17 U_{p}/\omega$ is much smaller than one explains our inability to observe an ATD plateau.
Physically, this small interval is due to the fact that the massive nuclei are unable to gain a substantial amount of energy in the field.
For a pulse with $I=8.9 \times 10^{13}$~$\mathrm{W/cm^{2}}$, it would require a wavelength of almost
3.7~$\mathrm{\mu m}$ to produce a plateau region with a large enough energy range to fit two KER peaks. 
        
Finally, the spectrum in Fig.~6 shows that high-order, non-linear processes are occurring:  the figure gives evidence for at least eleven photon absorption. 
Nevertheless, Fig.~4(b) suggests that the deviation from lowest-order perturbation theory is small.  
This apparent inconsistency can be reconciled by recognizing that the area under the $n=2$ peak dominates 
the energy integral needed to calculate dissociation probability from $dP/dE$. 
However, Fig.~6 also shows that the non-perturbative effects extend beyond the presence of high-order ATD peaks.
The diagonal line indicates $I^{n}$ behavior that matches the highest energy peaks as expected.
The lower peaks, however, do not meet this line, suggesting  
the contribution of high-order pathways in these peaks.  
Due to their dominating the energy integral of $dP/dE$, it is this deviation from $I^{n}$ that produces the non-perturbative behavior seen in Fig.~4(b) rather than the presence of the higher-order peaks.  

\subsection{Momentum Distribution}
\begin{figure}[tb!]
   \includegraphics[width=\columnwidth]{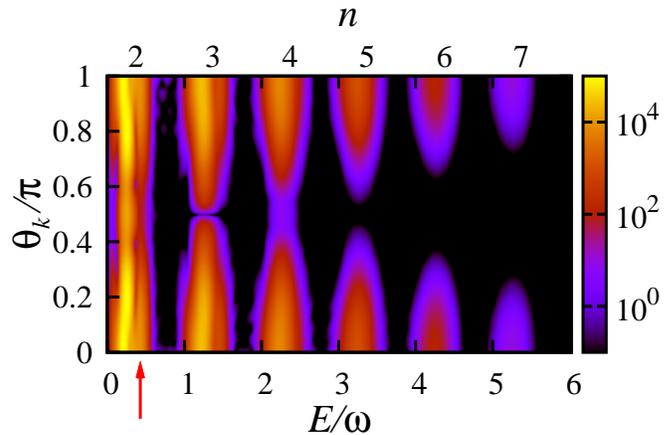} 
   \caption{(Color online.) ``Momentum'' distribution from Eq.~(\ref{MomDistDef}) plotted as a function of KER $E$ and angle $\theta_k$
           for an initial ro-vibrational state with $v$=3 and $J$=0 in a five cycle pulse with 
           an intensity of 8.9$\times$10$^{13}$~W/cm$^2$ and a wavelength of 2400~nm.  The 
           upper-axis label $n$ indicates the net number of photons absorbed.
   }
  \label{MomDistFig}
\end{figure} 
We can also use the atomic analogy to understand the distribution of relative nuclear momenta $\mathbf{k}$ 
\begin{align}
 \frac{\partial^{2} P}{\partial E \partial \theta_k} = 2\pi | \langle {\bf k} | \Psi(t_f) \rangle |^2 ,
 \label{MomDistDef}
\end{align}
where $\theta_{k}$ is the scattering angle with respect to the laser polarization and 
the state $|{\bf k}\rangle$ is an energy-normalized scattering state with asymptotic out-going momentum $\bf k$.
Note that Eq.~(\ref{KERDistDef}) is obtained from Eq.~(\ref{MomDistDef}) by integrating over the angle $\theta_k$.
The momentum distribution for $v=3$ and $\lambda=2400$~nm, the same case shown in Fig.~6, is shown in Fig.~7.
We, of course, still see the photon-spaced peaks with decreasing probability, as expected from the atomic ionization picture and Fig.~6. 
The angular distribution of each ATD peak is consistent with the identification of the number of photons involved.  
For instance, the peaks at $E/\omega=0.3,2.3,...$ are nonzero at $\theta_{k}/\pi =0.50$
as expected for even parity final states produced by the absorption of $n=2,4,...$ photons from the $J=0$ initial state. 
Similarly, the peaks at $E/\omega=1.3,3.3,...$ have a node at $\theta_{k}/\pi=0.50$
corresponding to an odd number
of photons absorbed --- specifically, $n=3,5,...$ --- from the $J=0$ initial state.

In both Fig.~\ref{KERDistFig} and Fig.~\ref{MomDistFig}, there is additional structure on the two-photon peak 
in the form of a peak at $E/\omega=0.50$ (indicated by arrows in the figures).
This peak can be explained by noting that the $v=3$ to $v=6$ transition is nearly on resonance for 2400 nm:
$\omega_{63}$=1.2$\omega$.  Even though the effective intensity at $\omega_{63}$ is reduced by seven orders
of magnitude, the resonant enhancement is still sufficient to produce a peak.
Moreover, this resonance enhanced multiphoton dissociation (REMPD) mechanism correctly predicts 
the peak to be at $E/\omega = 0.50$ in agreement with our calculations.
A REMPD peak is noticeable on the $n=3$ peaks in the figures as well.
For higher order peaks, however, the REMPD peak grows increasingly broad in
energy due to the bandwidth of the laser pulse, washing it out as a distinct
peak.  As its name suggests, REMPD is the analog of the well-known resonance 
enhanced multiphoton ionization in ATI spectra~\cite{Bransden}, and we expect it to be  
ubiquitous in long wavelength dissociation of $\mathrm{HeH^{+}}$ 
since these wavelengths are more likely to drive a bound-bound resonance.

\section{Summary and Outlook}

We have proposed $\mathrm{HeH^{+}}$ as a benchmark system for understanding the nuclear dynamics 
of heteronuclear molecules in intense laser fields.  However, given that dissociation of HeH$^+$
at the common intense laser wavelength of 800~nm is consistently
below useful experimental 
detection levels at intensities that will not also ionize the molecule,
it was necessary to explore other laser parameters.  We find that increasing the wavelength is an effective
way to increase the dissociation probability at modest intensities.  In particular, 
for wavelengths between 2000~nm and 2400~nm and intensities 
between 10$^{13}$~W/cm$^2$ and 10$^{14}$~W/cm$^2$, we see 
probabilities greater than one percent for $v=3$, 6, and 9, making dissociation of HeH$^+$ experimentally accessible.
In general, we expect even larger HeH$^+$ dissociation probabilities for wavelengths
longer than the 2400~nm considered here.

It turns out that HeH$^+$ is a rather unique molecule in that the minimum electronic excitation 
energy is relatively large.  While this property decreases its utility as a prototype for 
the behavior of heteronuclear molecules generally, it does allow for the surprising possibility
of understanding the nuclear dynamics of HeH$^+$ in terms of atomic ionization pictures.
The same reduction of the system to a single channel also has the intriguing consequence 
that all of the nuclear dynamics for a substantial range of laser parameters is induced
by the molecule's permanent dipole.  This lies in contrast to the electronic transition dominated
strong-field response of most every other molecule that has been studied to date.

So, while
HeH$^+$ may not be the general prototype we aimed for, it is still a very
interesting molecule to study.  
Even though much of the physics for this system can be understood using atomic pictures,
HeH$^{+}$ still exhibits molecular behavior that differentiates it from the atomic systems
that have been studied in the past.  As a molecule, for example, the initial
state of the system will actually be an incoherent distribution of ro-vibrational states
determined by the temperature and mechanism for creation of the molecule --- unlike atoms.
The strong-field dynamics of HeH$^+$ is thus not exactly molecule-like, and it is not exactly
atom-like.  But, with only two electrons, it {\em is} a benchmark system
for which precision calculations and experiments ought to be possible to elucidate
the dynamics of a system that displays physical behavior in 
limbo between that of molecule and that of an atom.

\begin{acknowledgments}
We are grateful for many useful discussions with I.~Ben-Itzhak, who 
provided the initial suggestion that we study $\mathrm{HeH^{+}}$ and 
gave much insight on its experimental accessibility.   We are also
grateful for many useful discussions with his group.
In addition, we wish to thank N.~Vaeck and the other authors of Ref.~\cite{0953-4075-43-6-065101}, 
who sent us their numerical data for the dipole matrix elements of $\mathrm{HeH^{+}}$. 
The authors acknowledge support 
the Chemical Sciences, Geosciences, and Biosciences Division, Office of Basic Energy Sciences, 
Office of Science, US Department of Energy.
DU further acknowledges support from the National Science Foundation (Grant Number PHY-0552878) during
the early stages of this work.
\end{acknowledgments}

\end{document}